\documentclass[conference]{IEEEtran}
\IEEEoverridecommandlockouts
\usepackage{cite}
\usepackage{amsmath,amssymb,amsfonts, amsthm, mathrsfs}
\usepackage{graphicx, float}
\usepackage{textcomp}
\usepackage{xcolor}
\usepackage{booktabs}
\usepackage{algorithm, algpseudocode}%
\usepackage[htt]{hyphenat}
\usepackage{mathtools}

\DeclarePairedDelimiter{\floor}{\lfloor}{\rfloor}
\PassOptionsToPackage{hyphens}{url}\usepackage{hyperref}

\newtheorem{remark}{Remark}
\allowdisplaybreaks[4]
\renewcommand{\baselinestretch}{0.89}
\def\BibTeX{{\rm B\kern-.05em{\sc i\kern-.025em b}\kern-.08em
    T\kern-.1667em\lower.7ex\hbox{E}\kern-.125emX}}
\begin{document}

\title{A Methodology for Quantifying Flexibility in a fleet of Diverse DERs$^\ast$
\thanks{$^\ast$This work was supported by the U.S. Department of Energy's ARPA-E award DE-AR0000694.}
}
\author{\IEEEauthorblockN{Adil Khurram \thanks{The authors are with the Department of Electrical and Biomedical Engineering, University of Vermont, Burlington, VT 05405 USA}, Luis A. Duffaut Espinosa and Mads R. Almassalkhi$^\dag$\thanks{$^\dag$M. Almassalkhi is co-founder of startup company Packetized Energy, which is commercializing aspects related to PEM.}}
\IEEEauthorblockA{\textit{Department of Electrical and Biomedical Engineering, University of Vermont,
Burlington, VT, USA} 
}}
\maketitle

\begin{abstract}
This paper addresses the question: how many distributed energy resources (DERs) are needed to provide $\pm1$MW of flexibility over a number of hours? For this purpose, a metric based on an ISO's own performance score is proposed. Then, a systematic procedure is presented and validated that makes use of either a simulator or the solution to an optimization problem based on a nominal analytical formulation to get flexibility in terms of kW-per-device. Furthermore, simulation-based analysis indicates that flexibility from different DER fleets adds linearly, that is, the total flexibility provided by a mixture of different DER types can be obtained as a convex combination of their individual kW-per-device flexibility. The proposed methodology is validated on ($i$) a centralized coordinator and ($ii$) a device driven DER coordination scheme called packetized energy management (PEM). Furthermore, the effect of heterogeneity as well as PEM specific parameters such as packet length and mean time-to-request on flexibility is also quantified.
\end{abstract}


\section{Introduction}
Distributed energy resources (DERs), such as smart electric water heaters (EWHs), and kW-scale energy storage systems (ESS) are inherently distributed, flexible and can be dispatched via \emph{demand dispatch} to provide services such as frequency regulation, peak-load reduction etc.~\cite{Brooks:2010gj, Mathieu:2013TPWRS}
However, in order to efficiently and reliably dispatch such DERs, utilities or grid operators are often faced with the question: how many DERs are necessary to provide $\pm1$MW of flexibility during some interval of time? This is not an easy question to answer given that DERs must also satisfy some desired level of operation based on human behavior.  

Demand dispatch coordinates DERs so that the utilities and grid operators can treat large DER populations as a single resource or specifically a virtual battery with the usual notion of state of charge (SoC) and energy capacity. By doing so, the operational details of individual DERs are abstracted away into a single charge or discharge command provided by the utility or grid operator and fulfilled by the demand dispatch scheme while satisfying quality of service (QoS) requirements. Virtual battery models have therefore been developed that can accurately capture the essential dynamics of the fleet~\cite{mathieu:2015TPWRS, Kundu:PSCC2018_geometric_flex, Hao-et-al_2015, Meyn:ACC2020_flex_TCLs, Barooah:ACC2020_flex_SD, Hughes;TPWRS2016_ident_VB_model,Soumya:acc2020}.
Flexibility is then defined as a set of constraints on SoC, aggregate power consumption and other related quantities in time domain~\cite{Meyn:ACC2020_flex_TCLs} and frequency domain~\cite{Barooah:ACC2020_flex_SD}. The authors in~\cite{mathieu:2015TPWRS} developed a time-varying virtual battery model for energy arbitrage. Geometric characterization of flexibility is presented in~\cite{Kundu:PSCC2018_geometric_flex} in which the aggregate flexibility of DERs is obtained as a Minkowski sum of individual DER flexibility. A virtual battery model is developed in~\cite{Hao-et-al_2015} for thermostatically controlled loads (TCL) that provides flexibility in terms of aggregate power and energy limits. 
A virtual battery model and population based models or macromodel have been developed for a device-driven demand dispatch scheme called packetized energy management (PEM) in~\cite{Almassalkhi:2018IMA,Duffaut-et-al_2019a,Duffaut-et-al_2018b,DuffautEspinosa:2020cdc}. In addition to analytical characterization of flexibility, data-driven identification of virtual battery models have also been proposed~\cite{Hughes;TPWRS2016_ident_VB_model, Soumya:acc2020}. 

\begin{figure}[htpb]
	\centering
	\vspace{-1em}
	\includegraphics[width=\columnwidth]{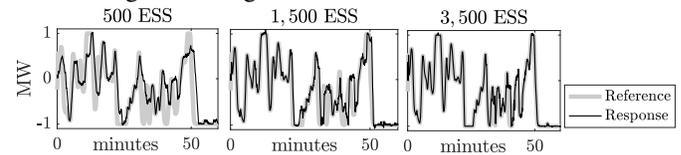}
	\vspace{-2em}
	\caption{PEM enabled ESS tracking a $1$-hour AGC signal scaled by $\pm1$MW.}
	\label{fig::motivation}
\end{figure}

Consider the aggregate response of a fleet of energy storage systems (ESS) shown in Fig.~\ref{fig::motivation} tracking a power reference signal under packetized energy management. As expected, the tracking error reduces as the size of the fleet increases and is projected to decrease even further if the number of devices increases more than $3,500$. However, from the point of view of the utility or grid operator, how many ESSs are sufficient for tracking such a signal?

The model-based methods~\cite{mathieu:2015TPWRS, Kundu:PSCC2018_geometric_flex, Hao-et-al_2015, Meyn:ACC2020_flex_TCLs, Barooah:ACC2020_flex_SD, Hughes;TPWRS2016_ident_VB_model} provide analytical bounds on power and energy of the virtual battery models. However, these bounds are either too conservative or obtained from steady-state analysis~\cite{DuffautEspinosa:2020cdc}. Therefore, this paper first introduces a notion of flexibility as the minimum number of DERs required to provide frequency regulation services such as automatic generation control (AGC) over a specified ($k$) number of hours and its reciprocal is defined as kW-per-device flexibility. To obtain kW-per-device, an iterative simulation based methodology is proposed that is agnostic to the demand dispatch scheme and only needs access to a simulator and historical AGC data. Next, the proposed methodology is validated on a fleet of DERs operating under a ($i$) centralized coordinator (CC) and ($ii$) PEM coordinator. Simulation results indicate that the flexibility of a mixture of different DER types may be obtained as a convex combination of their individual $k$-hour kW-per-device flexibility. Furthermore, it is shown through empirical evidence that as $k$ increases, the $k$-hour kW-per-device decreases and approaches the value obtained from steady-state analysis of population based models from~\cite{Duffaut-et-al_2019a}.

The paper is organized as follows. Section~\ref{sec::PEM_prelim} describes the centralized and PEM coordinator. The methodology to obtain $\pm1$ MW flexibility is presented in section~\ref{sec::get_flex} which is applied to DERs of the same type and to a mixture of diverse DERs in section~\ref{sec::flex_DERs}. In section~\ref{sec::flex_in_ss}, the population based models are used to provide steady-state flexibility and section~\ref{sec::conclusion} concludes the paper.

 

\section{DER coordination schemes} \label{sec::PEM_prelim}
This section describes a centralized DER coordination scheme as well as a distributed device-driven coordination scheme called PEM. The main difference between the centralized coordinator (CC) and the PEM coordinator is that the CC has access to full information about energy states and operating modes of all DERs in the fleet but provides no QoS guarantees whereas PEM is a device driven scheme in which a DER's request to consume power is driven by their energy states and guarantees QoS by allowing DERs to temporarily opt-out of PEM.

\subsection{Centralized DER coordination}

Let $x[k]$ be the energy state of the DER under the centralized coordinator, $x_{\text{set}}$ be the set-point, $[\underline{x}, \overline{x}]$ be the operating dead-band where $\underline{x}$ and $\overline{x}$ are the lower and upper energy limits and $P_{\text{dem}}$ be total power consumption of the fleet. At any time $k$, a DER can ($i$) consume power from the grid in charge mode, ($ii$) inject power into the grid in discharge mode or ($iii$) be in standby mode in which the DER is neither consuming nor injecting power into the grid. Furthermore, the CC transmits control commands to each DER at regular intervals, that instructs the DERs to either continue in its current mode or switch to the instructed mode as shown in Fig.~\ref{fig::PEMschematic}. CC determines control commands by prioritizing DERs depending upon their energy states $x[k]$ as explained next.

Given a power reference signal $P_{\text{ref}}$, the central coordinator determines the DERs whose operating state needs to be changed so that the tracking error is minimal, in the following manner. Let $e[k] = P_{\text{ref}}[k] - P_{\text{dem}}[k]$ be the tracking error at time $k$. If $e[k] > 0$ then the CC first starts with the DERs in standby mode and sends a command signal instructing the DERs to turn ON and start charging. Priority is given to those DERs whose $x[k]$ is lower. If the number of DERs in standby mode are not sufficient to drive the $e[k]$ to zero, then the CC instructs DERs in discharge mode to stop injecting power into the grid and transition to standby, again prioritizing DERs with lower $x[k]$. Similarly, if $e[k] < 0$, then the CC first switches DERs in standby mode to discharge mode followed by instructing the DERs in charge mode to turn off. Here, priority is given to those DERs with higher $x[k]$.

\subsection{Distributed DER coordination (with PEM)}
\begin{figure}[htpb]
	\centering
	\vspace{-1em}
	\includegraphics[width=\columnwidth]{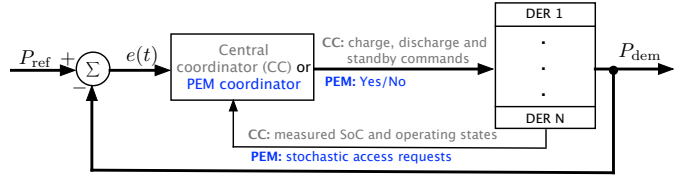}
	\vspace{-1.5em}
	\caption{Closed loop feedback system under the centralized coordinator (with corresponding inputs/outputs shown in grey) and the PEM coordinator (with corresponding inputs/outputs shown in blue), tracking the reference $P_{\text{ref}}$ and the aggregate power consumption is given by $P_{\text{dem}}$.}
	\label{fig::PEMschematic}
	\vspace{-1em}
\end{figure}

To illustrate a distributed DER coordination scheme, consider PEM, which has been validated in simulation for diverse DERs in~\cite{Almassalkhi:2018IMA} and systematically characterized in~\cite{Duffaut-et-al_2019a,Duffaut-et-al_2018b, DuffautEspinosa:2020cdc}. A brief description of PEM is presented here.
DERs operating under PEM~\cite{Almassalkhi:2018IMA} can be in one of the four logical modes, ($i$) charge, ($ii$) discharge, ($iii$) standby and ($iv$) opt-out. The PEM scheme is summarized as follows:
\begin{itemize}
    \item[$i.$] At any time $k$, a DER measures its energy state $x[k]$.
    \item[$ii.$] If $x \in [\underline{x}, \overline{x}]$ then the DER makes either a charge request or a discharge request but not both. The probability of making a charge request is given by,
    \begin{align}
        P_{\mu}(x[k]) &= 1 - e^{-\mu(x[k])\Delta t}\label{eq:PEM_stoch_req},\\
        \mu_{\text{c}}(x[k]) &= \left\{\begin{matrix}
        0, & \text{if } x[k]\ge \overline{x}\\
        m_R(\frac{\overline{x} - x[k]}{x[k] - \underline{x}}) (\frac{x_{\text{set}} - \underline{x}}{\overline{x} - x_{\text{set}}}), & \text{if } x[k]\in (\underline{x}, \overline{x})\\
        \infty, & \text{if } x[k]\le \underline{z} 
        \end{matrix}\right. \nonumber
    \end{align}
    where, $\Delta t$ is the time-step, $x_{\text{set}}$ is the set-point and $m_R$ is called as the mean time-to-request (MTTR). Discharge request probability can be obtained in a similar manner. If the charge (discharge) request is accepted, then the DER transitions to charge (discharge) mode and consumes (discharges) power from (into) the grid for a pre-specified time called \emph{packet length} that begets an energy packet. Once the packet has expired, DERs moves back to standby mode.
    \item[$iii.$] If $x \notin [\underline{x}, \overline{x}]$ then the DER opts-out of PEM and either charges (if $x[k] < \underline{x}$) or discharge (if $x[k] > \overline{x}$) until $x[k]$ returns within $[\underline{x}, \overline{x}]$ and transitions back to standby mode.
\end{itemize}
The probability of request in~(\ref{eq:PEM_stoch_req}) has been designed so that the DERs with low $x[k]$ request to charge more frequently than those with higher $x[k]$. The closed loop PEM system is shown in~Fig. \ref{fig::PEMschematic} where $P_{\text{ref}}$ is the reference signal and $P_{\text{dem}}$ is the aggregate power consumption of the fleet.
\section{Methodology to obtain flexibility} \label{sec::get_flex}
This section describes the procedure to obtain kW-per-device flexibility that can be generalized to any type of load coordination scheme that may or may not guarantee QoS using the performance metrics presented next.
\subsection{PJM performance scoring}
Pennsylvania, Jersey, Maryland Power Pool (PJM) is a regional transmission organization that coordinates the movement of wholesale electricity and is part of the Eastern interconnection in the United States. PJM measures the performance of a resource, which in this case are DERs under PEM or CC,
providing ancillary services to the grid using an average of three metrics, ($i$) \emph{Accuracy} ($x_{\text{a}}$), ($ii$) \emph{Delay} ($x_{\text{d}}$) and ($iii$) \emph{Precision} ($x_{\text{p}}$), called \emph{Composite} score ($x_{\text{c}}$) described below:
\begin{itemize}
    \item[$i.$] \emph{Accuracy} score ($x_{\text{a}}$) is measured using the correlation of the regulation signal with the response of flexible resources over a $5$ minute rolling window and a sampling interval of $10$ seconds. The maximum correlation over each of the $5$ minute rolling windows is averaged to obtain the accuracy for each hour.
    \begin{align*}
    x_{\text{a}} =& {\text{max}}_{t_j = 0:10:300 {\text{sec}}}\{{\text{corr}}(P_{\text{ref}}(t_0 : t_0 + 3600), \\
    & P_{\text{dem}}(t_0 + t_j : t_0 + t_j + 3600)) \}
    \end{align*}
    \item[$ii.$] \emph{Delay} score ($x_{\text{d}}$) is measured using the point of highest correlation between the regulation signal and the response and is defined as $t_k = {\text{argmax}}_{t_j} ({\text{corr}}(P_{\text{ref}})$ which results in, 
    \begin{align*}
        x_{\text{d}} = {\text{max}}\left\{ 1, \left|\frac{t_k - 10 - 300}{300} \right| \right\}
    \end{align*}
    \item[$iii.$] \emph{Precision} score ($x_{\text{p}}$) is the instantaneous error between the regulation signal and the response, 
    \begin{align*}
        x_{\text{p}} = 1-\text{average} \left\|\frac{P_{\text{ref}} -P_{\text{dem}}}{P_{\text{ref}}}\right\|.
    \end{align*}
\end{itemize}

\begin{figure}[htpb]
	\centering
	\includegraphics[width=\columnwidth]{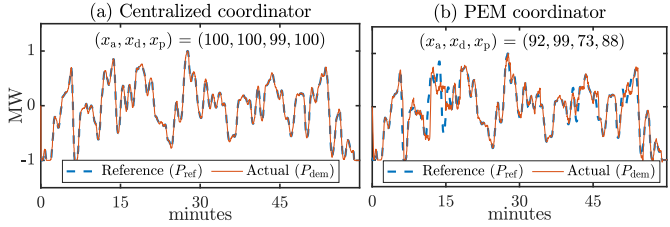}
	\vspace{-1.5em}
	\caption{PJM's \emph{Accuracy}, \emph{Delay}, \emph{Precision} and \emph{Composite}, $(x_{\text{a}}, x_{\text{d}}, x_{\text{p}}, x_{\text{c}})$, scores are shown for a fleet of $1,500$ ESS tracking a  $1$-hour AGC signal scaled by $1$MW under CC and PEM are plotted in sub-figures ($a$) and ($b$) respectively. CC achieves almost perfect scores whereas PEM achieves high \emph{Accuracy} and \emph{Delay} score but comparatively lower \emph{Precision} score.}
	\label{fig::pjm_score_example}
	\vspace{-1em}
\end{figure}

It should be mentioned here for a time period equal to $k$-hours, these metrics are calculated over $50$ minute rolling time windows resulting in $n_k = \floor*{\frac{6k-1}{4}}$ values for each of \emph{Accuracy}, \emph{Delay} and \emph{Precision scores}, where $\floor*{.}$ is the floor function. Therefore, $x_{\text{u}}$ for all ${\text{u}}\in\{\text{a,\, d,\, p} \}$ is obtained as the minimum of $n_k$ scores. Then, the \emph{Composite} score is determined as $x_{\text{c}} = \frac{1}{3} \left( x_{\text{a}} + x_{\text{d}} + x_{\text{p}} \right)$. Fig.~\ref{fig::pjm_score_example} shows $1,500$ ESS tracking an AGC signal scaled by $1$MW over $1$-hour under both coordinators presented in the previous section along with the performance metrics. The centralized coordinator in Fig.~\ref{fig::pjm_score_example}(a) achieves almost perfect scores for all three metrics since it assumes full knowledge and control over ESS's operating state. On the other hand, Fig.~\ref{fig::pjm_score_example}(b) shows that under PEM with $2$ minute packet-length and mean time-to-request, the fleet of ESS achieves comparatively higher \emph{Accuracy} and \emph{Delay} scores than \emph{Precision} score. Hence, kW-per-device flexibility in the next section is based on \emph{Precision} score only.

\subsection{Flexibility over $k$-hours}
The flexibility is defined in terms of the minimum number of DERs ($N_{\text{DER}}^{\text{min}}$) needed to track an AGC signal scaled by $1$MW over $k$-hours and achieve a desirable performance score. The kW-per-device ($\zeta_{\text{DER}}^k$) is then obtained as $\zeta_{\text{DER}}^k = 1,000 (N_{\text{DER}}^{\text{min}})^{-1}$kW, where $k$ is the number of hours of the AGC signal. Since PEM system achieves high accuracy and delay scores, therefore, only \emph{Precision} score is used as a metric as explained next.

\subsection{Procedure to obtain flexibility}
The procedure to obtain flexibility uses historical AGC data which may span over several years. The main idea behind this method is to select $m$ number of $k$-hour AGC signals instead of testing over the entire data-set and use a simulator to obtain $N_{\text{DER}}^{\text{min}}$ as described by Algorithm~\ref{alg:procedure_Nmin_der}, similar to~\cite{Soumya:acc2020}. A Matlab based PEM simulator is used for this purpose. Furthermore, the set of chosen AGC signals should be representative of the historical data-set and the corresponding selection criterion is presented in the next section.

Let $\xi_{\text{DER}}^{\text{param}}$ denote the parameter set corresponding to the DER under consideration and let the $k$-hour AGC signal scaled by $1$MW be denoted by $\vec{P}^{k}_{{\text{agc}}, i} \in \mathbb{R}^{K}, K = 3600 k (\Delta t)^{-1}$,
where $\Delta t$ is the time resolution in seconds of the AGC signal. Then, $\{\vec{P}^{k}_{{\text{agc}}, i}\}\, \forall \, i = 1, \dots, m$ is the set of representative $k$-hour AGC signals. As described in Algorithm~\ref{alg:procedure_Nmin_der}, starting from an initial fleet size $N_0 = N_{\text{start}}$, the \texttt{simulator} is used to track each of $\vec{P}^{k}_{{\text{agc}}, i}$ over $k$-hours and
\emph{Precision} score $x^0_{{\text{p}}, i}$ is calculated. If $x^0_{{\text{p}}, i} \ge x_{\text{p,des}} \, \forall \, i = 1, \dots, m$ where $x_{\text{p,des}}$ is the desired \emph{Precision}, then $N_0$ is sufficient to provide $k$-hour $1$MW flexibility and $N_{\text{DER}}^{\text{min}} = N_0$, otherwise, the process continues with $N_1 = N_0 + \Delta N_{\text{DER}}$ where $\Delta N_{\text{DER}}$ is the step-size for population size, until 
\begin{align}
    x_{{\text{p}}, i}^j > x_{\text{p,des}} \, \, \forall \, i = 1, \dots, m,  \label{eq:stopping_criterion}  
\end{align}
where $j$ is the iteration number. Finally, kW-per-device ($\zeta^{k}_{\text{DER}}$) is obtained by $\zeta^{k}_{\text{DER}} = 1,000 (N^{\text{min}}_{\text{DER}})^{-1}$kW. 
\vspace{-0.5em}

\begin{algorithm}[htbp]
  \caption{Procedure to obtain $N_{\text{DER}}^{\text{min}}$}
    \hspace*{\algorithmicindent} \textbf{Input}: $\{ P_{\text{agc},i}^k \}_{i=1}^m,\,N_{\text{start}},\, \Delta N_{\text{DER}},\, \xi^{\text{param}}_{\text{DER}}$ \\
    \hspace*{\algorithmicindent} \textbf{Output}: $N_{\text{DER}}^{\text{min}}$
  \begin{algorithmic}[1]
    \For{$i \gets 1, m$}\Comment{For each selected $k$-hour $\vec{P}_{\text{agc}, i}^k$}
        \State $N_0 \gets N_{\text{start}}$
        \State $j \gets 1$
        \While{$x_{\text{p}, i}^j \le x_{\text{p,des}}$} \Comment{Stopping criterion~(\ref{eq:stopping_criterion})}
            \State $N_j \gets N_{j-1} + \Delta N_{\text{DER}}$
            \State $x_{\text{p}, i}^j \gets \texttt{simulator}(N_j, \vec{P}_{\text{agc}, i}^k, \xi^{\text{param}}_{\text{DER}}$)
            \State $j \gets j + 1$
        \EndWhile
        \State $N_{\text{DER}, i}^{\text{min}} \gets N_j$
    \EndFor
    \State \textbf{return} $N_{\text{DER}}^{\text{min}} \gets  \max \{ N_{\text{DER}, i}^{\text{min}}\}_1^m$
  \end{algorithmic}
  \label{alg:procedure_Nmin_der}
\end{algorithm}

\vspace{-0.8em}  
\subsection{Statistics of AGC and selection criterion}
In order to obtain flexibility using the proposed methodology, the selection of $\vec{P}^{k}_{{\text{agc}},i}$ should be such that it is representative of the AGC signal. For example,~\cite{Soumya:acc2020} uses $200$ two-hour periods of the AGC signal to identify a virtual battery model. This work, however, focuses on selecting AGC signal based on hourly mean since it is reasonable to capture flexibility. Ongoing work is studying different metrics such as entropy.

The historical AGC data used in this work is obtained from~\cite{PJM_AGC_url} which is normalized in the range $[-1, 1]$ with $2$ second resolution ($\Delta t = 2$) and spans over a full year between July $2018$ and June $2019$. PJM has implemented a conditional neutrality controller that generates the regulation signal that is energy neutral over longer time periods such as a full day, meaning that its mean value is zero~\cite{PJM_energy_neutral}. However, within a span of few hours, the mean is non-zero that causes the population to either charge or discharge as an aggregate when providing services to the grid.

\begin{figure}[htbp]
	\centering
	\includegraphics[width=1\columnwidth]{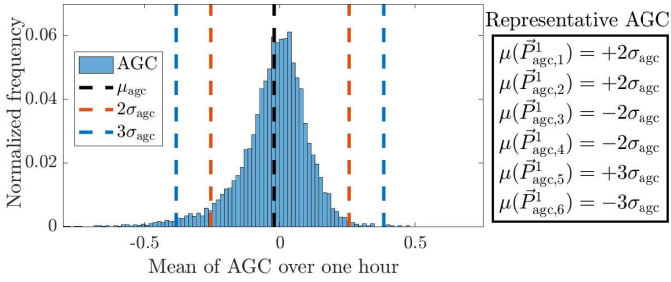}
	\vspace{-2em}
	\caption{Distribution of $1$-hour mean values of the AGC over a full year between July $2018$ and June $2019$. Red dotted line shows $\pm 2\sigma_{\text{agc}}$ and blue dotted line is the $\pm 3 \sigma_{\text{agc}}$ where $\sigma_{\text{agc}}$ is the standard deviation of the distribution. The mean of the chosen $\vec{P}_{\text{agc}, i}^i$, $i = 1, \dots 6$ are also shown.}
	\label{fig::agc_hourly_mean}
	\vspace{-1em}
\end{figure}

Consider first the mean values of the AGC signal calculated over one hour ($k=1$) as shown in Fig.~\ref{fig::agc_hourly_mean}. It can be seen from Fig.~\ref{fig::agc_hourly_mean} that the distribution is biased towards the left with $\mu_{\text{agc}} = -0.021$ and the standard deviation $\sigma_{\text{agc}} = 0.272$. 
The distribution in Fig.~\ref{fig::agc_hourly_mean} resembles a normal distribution for which approximately $99\%$ of the data lies in the interval $\mathbb{I}_{\text{agc}}:=[-3\sigma_{\text{agc}}, +3\sigma_{\text{agc}}]$. Therefore, for all the simulation studies presented in this paper, six representative $k$-hour AGC signals are randomly selected (that is, $m=6$) so that their mean lies in the interval $\mathbb{I}_{\text{agc}}$. Specifically, two of $\vec{P}^{k}_{{\text{agc}},i}$ have mean equal to $+2\sigma_{\text{agc}}$, two have mean equal to $-2\sigma_{\text{agc}}$ and the last two have mean equal to $+3\sigma_{\text{agc}}$ and $-3\sigma_{\text{agc}}$ respectively. Finally, for $k>1$, each of the selected $\vec{P}^{1}_{{\text{agc}},i}$ are repeated over $k$-hours to get $\vec{P}^{k}_{{\text{agc}},i}$.

\section{Flexibility of DERs}\label{sec::flex_DERs}
The procedure developed in the previous section that consists of selecting representative AGC signals and Algorithm~\ref{alg:procedure_Nmin_der}, is applied to obtain flexibility of a fleet of ESSs and EWHs. In this section, the focus is on $1$-hour and $\pm1$MW flexibility.
\subsection{Flexibility of ESS}
Consider a fleet of ESS with parameter set $\xi^{\text{param}}_{\text{ESS}} = \{ P_{\text{rate}}^{\text{c}}, P_{\text{rate}}^{\text{d}}, \eta_{\text{c}}, \eta_{\text{d}}, E_{\text{cap}}, x_{\text{set}}, \underline{x}, \overline{x} \} $ and the corresponding values given in Table~\ref{table::DER_param}. For the centralized coordinator, denoted with the sub-script CC in this section, the $1$-hour flexibility ($k=1$) is obtained in terms of kW-per-device ($\zeta_{\text{ESS,CC}}^{1}$) by first selecting $m=6$, representative AGC signals $\vec{P}^{k}_{{\text{agc}},i}$. Next, the desired \emph{Precision} score is set to be $x_{\text{p,des}} = 70\%$, the initial number of ESS to $N_{\text{start,CC}} = 50$ and $\Delta N_{\text{ESS,CC}} = 50$. Application of Algorithm~\ref{alg:procedure_Nmin_der} then  results in $N_{\text{ESS,CC}}^{\text{min}} = 200$ that translates to $\zeta_{\text{ESS,CC}}^{1} = 5$kW which is the maximum flexibility that can be obtained from an ESS rated at $5$kW. It should be mentioned here that $5$kW-per-ESS corresponds to $1$-hour $1$MW AGC signal only.

\begin{figure}[htbp]
	\centering
	\includegraphics[width=1\columnwidth]{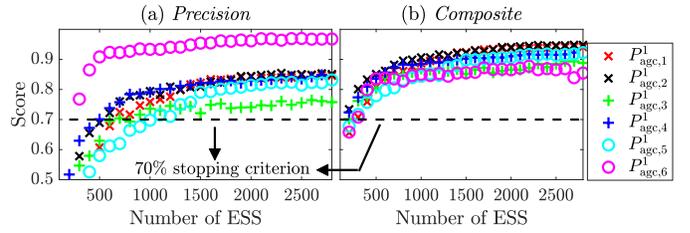}
	\vspace{-1.5em}
	\caption{For each of the six representative $1$-hour AGC signals, the number of ESS is varied from $N_{\text{start,PEM}} = 100$ to $3,000$ (for illustration purposes), and the \emph{Precision} as well as \emph{Composite} score is plotted in sub-figures (a) and (b) respectively. The $N_{\text{ESS}}^{\text{min}}$ required to satisfy stopping criterion~(\ref{eq:stopping_criterion}) is $1,100$ which translates to $\zeta_{\text{ESS,PEM}}^{1} = 0.91$kW per ESS of flexibility.}
	\label{fig::912_ess_flex}
	\vspace{-1em}
\end{figure}

Similarly for the PEM coordinator, Algorithm~\ref{alg:procedure_Nmin_der} is applied with the selected AGC signals, $N_{\text{start,PEM}} = 100$ and $\Delta N_{\text{ESS,PEM}} = 200$. Fig.~\ref{fig::912_ess_flex} shows that the precision score ($x^j_{{\text{p}},i}$) increases as $N_j$ increases, however, $N_{\text{ESS,PEM}}^{\text{min}} = 1,100$ ESS are sufficient to satisfy the stopping criterion~(\ref{eq:stopping_criterion}) resulting in $\zeta_{\text{ESS,PEM}}^{1} = 0.91$kW. For the purpose of illustration, both \emph{Precision} scores and \emph{Composite} scores are plotted in Fig.~\ref{fig::912_ess_flex}, sub-figures ($a$) and ($b$) respectively, for $N_j > N_{\text{ESS,PEM}}^{\text{min}}$ that shows good performance. 

Comparing kW-per-device for both coordinators, it is obvious that $\zeta_{\text{ESS,CC}}^{1} > \zeta_{\text{ESS,PEM}}^{1}$ strictly under the definition of flexibility considered in this work. However, CC requires that a large amount of data be streamed regularly between CC and DERs which includes operating states, SoC/power measurements and command signals for DERs. This type of controller is suitable for small populations and becomes impractical when deployed to fleets consisting of thousands of DERs mainly due its substantial bandwidth and computation requirements. Distributed coordinators, on the other hand, are more suitable for such cases since communication between the coordinator and DERs is designed to be minimal. For example in PEM, coordination is achieved via the device-driven request-response mechanism. Therefore, in the following sections, the focus is on PEM coordinator and the sub-script PEM is dropped henceforth.

\begin{figure}[htbp]
	\centering
	\vspace{-1em}
	\includegraphics[width=1\columnwidth]{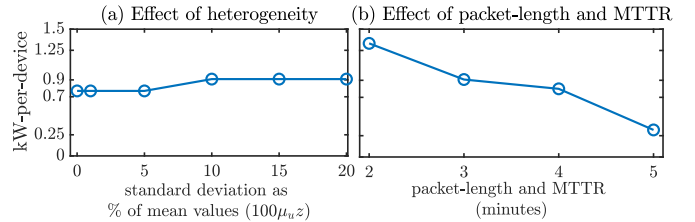}
	\vspace{-1em}
	\caption{Sub-figure (a) shows that heterogeneity has a small effect on kW-per-device whereas in (b), the increase in packet length and MTTR (in minutes) reduces kW-per-device for ESS.}
	\label{fig::kW_per_ESS_vs_PKT_MTTR}
	\vspace{-1em}
\end{figure}

In the following remarks, the effect of parameter heterogeneity is first investigated in Fig.~\ref{fig::kW_per_ESS_vs_PKT_MTTR}(a) and is found to have a small impact on kW-per-device. This allows meaningful analysis of the effect of PEM-specific parameters on flexibility as shown in Fig.~\ref{fig::kW_per_ESS_vs_PKT_MTTR}(b) and highlighted in Remark~\ref{rem:pl_mttr}.


\begin{remark} \label{rem:htr}
    The effect of heterogeneity is studied in simulations in which each of the parameters $y \in \xi_{\text{ESS}}^{\text{param}}$ is drawn from an uncorrelated normal distribution $\mathcal{N}(\mu_y, \sigma_y)$ with mean equal to the corresponding parameters in Table~\ref{table::DER_param} and the standard deviation $\sigma_y$ is set as the $z \%$ of the mean values, that is, $\sigma_y = z \mu_y $, $z \in [0, 1]$. A small increase in kW-per-device is observed in Fig.~\ref{fig::kW_per_ESS_vs_PKT_MTTR}(a) from $0.77$kW to $0.91$kW as $z$ increases. Future work will focus on quantifying this effect.
\end{remark}
\begin{remark}\label{rem:pl_mttr}
The kW-per-device ($\zeta_{\text{ESS}}^{1}$) flexibility decreases with the increase in packet-length and MTTR as shown in Fig.~\ref{fig::kW_per_ESS_vs_PKT_MTTR}(b). Flexibility provided by ESS with $2$~minute packet-length and MTTR is about $1.25$kW whereas $5$~minute packet-length and MTTR reduces the kW-per-device to $0.25$kW. The reason is that shorter packet-length/MTTR allows tighter tracking of the AGC signal resulting in a better Precision score. However, it should be noted that although shorter packet-length/MTTR improves flexibility but at the cost of higher communication between PEM coordinator and DERs. Work is ongoing to characterize the trade-offs between flexibility and communication overhead.
\end{remark}

\subsection{Flexibility of EWHs}
\begin{figure}[t]
	\centering
	\includegraphics[width=\columnwidth]{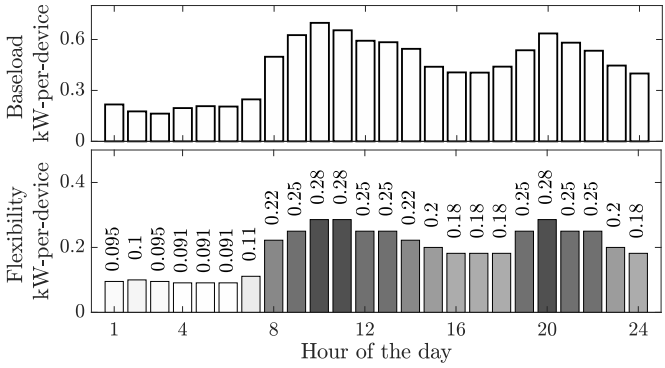}
	\vspace{-1.8em}
	\caption{The average end-use consumption (top) affects the available flexibility of the fleet (bottom). The bottom plot shows the $1$-hour flexibility provided by an EWH. The flexibility is obtained from simulations over the chosen set of hourly AGC signals so that the stopping criterion~(\ref{eq:stopping_criterion}) is satisfied. The kW-per-device is greater in the morning (between $8$am and $2$pm) and evenings (between $8$pm and $11$pm) which is due to larger end-use consumption during these times as plotted at the top. }
	\label{fig::912_ewh_flex}
	\vspace{-1em}
\end{figure}

\begin{figure}[t]
	\centering
	\includegraphics[width=\columnwidth]{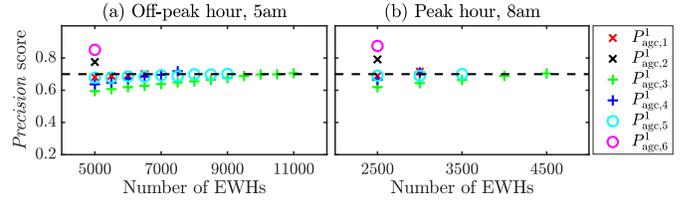}
	\vspace{-1em}
	\caption{PJM \emph{Precision} scores are shown here for different number of EWHs between $5$am and $6$am in sub-figure (a) which represents the time of day with low end-use consumption, that is, off-peak hour and requires at least $11,000$ EWHs to provide $1$MW of flexibility. The peak-hour between $8$am and $9$am is shown in sub-figure (b) for which only $4,100$ EWHs are sufficient to provide $1$MW flexibility.
	}
	\label{fig::912_agc_5hr_pjm_score}
	\vspace{-1.5em}
\end{figure}

Similar to ESS, kW-per-device flexibility of electric water  heaters (EWHs) during $1$-hour is obtained using the Matlab based simulator and Algorithm~\ref{alg:procedure_Nmin_der}. The parameters of EWHs are given by the set $\xi_{\text{EWH}}^{\text{param}} = (P^\text{c}_{\text{rate}}, L, x_{\text{set}}, \underline{x}, \overline{x}, x_{\text{amb}})$ and their values are in Table~\ref{table::DER_param}. It should be noted here that the end-use consumption of EWHs varies throughout the day~\cite{Khurram:2020epsr}. As a result, the nominal power consumption or baseload also changes as shown in the top plot of Fig.~\ref{fig::912_ewh_flex} and the flexibility also differs depending upon the hour of the day. The kW-per-device ($\zeta_{\text{EWH}}^{1}$) is, therefore, computed for each of the $24$ hours and is shown at the bottom plot of Fig.~\ref{fig::912_ewh_flex}. The $\zeta_{\text{EWH}}^{1}$ is larger during peak hours (e.g. between $8$am and $11$am) when the end-use consumption is higher than the off-peak hours (e.g. between $3$pm and $5$pm) when the end-use consumption is comparatively lower. This difference is because higher end-use consumption increases the need of EWHs in standby to consume energy that produces more requests in PEM and results in higher upward flexibility. On average, during peak hours, an EWH can provide $0.25$kW and during off-peak hours, it reduces to about $0.1$kW. The reason is that during off-peak hours, for example the hour starting at $5$am in Fig.~\ref{fig::912_agc_5hr_pjm_score}(a), the baseload is less than $1$MW for the fleet size less than $5,000$. $N_{\text{start}} = 5,0000$ in Algorithm~\ref{alg:procedure_Nmin_der}. On the other hand, for the peak hour starting at $8$am in Fig.~\ref{fig::912_agc_5hr_pjm_score}(b), $N_{\text{start}} = 2,500$ for which the baseload is greater then $1$MW.

\begin{table}[t]
\centering
\caption{DER parameters}
\label{table::DER_param}
\begin{tabular}{ l c l c}
\toprule
$\xi^{\text{param}}_{\text{EWH}}$& Value &$\xi^{\text{param}}_{\text{ESS}}$ & Value\\
\midrule
$P^{\text{c}}_{\text{rate}}$ &  $4$kW & $P^{\text{c}}_{\text{rate}}$ = $P^{\text{d}}_{\text{rate}}$ & $5$kW\\
$L$& $303$ liters & $\eta_{\text{c}} = \eta_{\text{d}}$ & $95\%$ \\
$x_{\text{amb}}$& $70^{\circ}$F & $E_{\text{cap}}$ & $13.5$kWh \\
$x_{\text{set}}$& $130^{\circ}$F &$x_{\text{set}}$ & $50\%$ \\
$[\underline{x}, \overline{x}]$  & $[120,\,140]^{\circ}$F & $[\underline{x}, \overline{x}]$  & $[10,\,90] \% $\\
\bottomrule
\vspace{-3em}
\end{tabular}
\end{table}

\subsection{Flexibility of a mixture of diverse DERs}
\begin{figure}[htbp]
	\centering
	\includegraphics[width=\columnwidth]{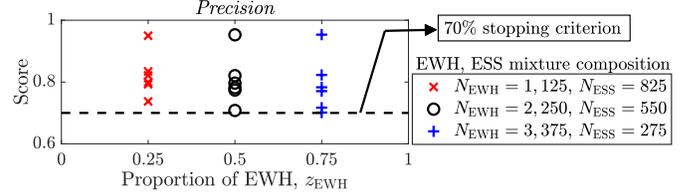}
	\vspace{-0.5em}
	\caption{To achieve the same $1$ MW of flexibility using a mixture of EWHs and ESSs, the kW-per-device flexibility is used to determine the composition of the fleet. Three cases are considered here, (i) $25\%$ EWHs, $75\%$ ESSs, (ii) $50\%$ EWHs, $50\%$ ESSs, (iii) $75\%$ EWHs, $25\%$ ESSs. All three cases result in \emph{Precision} score greater than $70\%$. However, it should be noted that increasing the percentage of ESS in the mixture results in better \emph{Precision}. The reason is attributed to the fact that ESS can discharge (that is, to request a discharge packet in PEM) and provides greater downward flexibility.}
	\label{fig::912_ewh_ess_8hr}
	\vspace{-1.5em}
\end{figure}

In this section, the relation between kW-per-device flexibility of the same DER type and that of a mixture of diverse DERs is studied. Assuming that $\zeta_{\text{ESS}}^1$ and $\zeta_{\text{EWH}}^1$ obtained in the previous section extends to a mixture of EWH and ESS, a diverse fleet consisting of the number of ESS given by $N_{\text{ESS}} = 1,000z_{\text{ESS}}(\xi_{\text{ESS}}^1)^{-1}$ and the number of EWHs given by $N_{\text{EWH}} = 1,000z_{\text{EWH}}(\xi_{\text{EWH}}^1)^{-1}$ is considered where $z_{\text{ESS}}, z_{\text{EWH}} \in [0,1]$ is the proportion of ESS and EWH in the mixture respectively. Three cases are studied, with EWH and ESS proportions ($i$) $25\%$ EWH, $75\%$ ESS, ($ii$) $50\%$ EWH, $50\%$ ESS, (iii) $75\%$ EWH, $25\%$ ESS. Fig.~\ref{fig::912_ewh_ess_8hr} shows the performance scores for the peak hour starting at $8$am and evaluated for each of six chosen $1$-hour AGC signals. All three cases result in \emph{Precision} score greater than $70\%$ indicating that the kW-per-device flexibility of a mixture may be obtained as a convex combination of individual flexibility. However, in Fig.~\ref{fig::912_ewh_ess_8hr}, as the proportion of EWH increases in the mixture, the \emph{Precision} score decreases. This is because ESS can provide downward flexibility from discharging packets resulting in better \emph{Precision}.


\section{Long term flexibility} \label{sec::flex_in_ss}
In this section, simulation-based analysis suggests that 
the $k$-hour kW-per-device flexibility obtained using Algorithm~\ref{alg:procedure_Nmin_der} approaches that from a PEM macromodel as $k$ increases. The focus is on ESS but the results can be applied to other DERs.
\subsection{Macromodel for PEM}
State bin transition model or macromodel is a Markovian model that has previously been developed by the authors and captures the aggregate behavior of a fleet of PEM enabled DERs~\cite{Duffaut-et-al_2019a, Duffaut-et-al_2018b}. Steady state statistics of the macromodel are used in this section to compute the long term flexibility of DERs. A brief description of macromodel is presented next.

Let $ \tilde{x} \in \mathbb{R}^{n_{\text{b}}}$ be the vector of states obtained by partitioning the operating dead-band of DERs $[\underline{x}, \overline{x}]$ into $n_{\text{b}}$ bins. The complete state space of the macromodel is then given by $q = (q_{\text{chg}}, q_{\text{dis}}, q_{\text{sb}}, q_{\text{opt}})^\top$, where $q \in \mathbb{R}^{4n_{\text{b}}}$ and $q_u$ are identical copies of $\tilde{x}$ for all  $u \in \{ \text{chg}, \text{dis}, \text{sb}, \text{opt}\}$. Furthermore, the aggregate dynamics of the PEM system evolves according to,
\begin{align}
    q[k+1] = f(\beta[k], \beta^-[k], q[k]),\,\, P_{\text{dem}}[k] = h(q[k]) \label{eq:PEM_nonlin_dyn}
\end{align}
where, $\beta = (\beta_{\text{chg}}, \beta_{\text{dis}})^\top$, $\beta^- = (\beta^-_{\text{chg}}, \beta^-_{\text{dis}})^\top$, $f$ is a non-linear mapping, $f:\mathbb{R}^{4+4n_{\text{b}}} \rightarrow \mathbb{R}^{4n_{\text{b}}}$ and $h$ is a linear map $h:\mathbb{R}^{4n_{\text{b}}} \rightarrow \mathbb{R}$ to get $P_{\text{dem}}$ . Here $\beta_{\text{chg}} (\beta_{\text{dis}}$) is the proportion of charging (discharging) requests accepted by the PEM coordinator out of the total DERs in standby mode. Similarly, $\beta^-_{\text{chg}} ( \beta^-_{\text{dis}}$) is the proportion of charging (discharging) DERs that have completed their energy packet and now transition from  charge (discharge) mode to standby mode. These transitions are defined in the non-linear map $f(.)$ and the reader is referred to~\cite{Duffaut-et-al_2019a,Duffaut-et-al_2018b} for further details as well as the exact description of the macromodel.
\subsection{Multiple hour flexibility} 
\begin{figure}[htbp]
    \vspace{-0.5em}
	\centering
	\includegraphics[width=\columnwidth]{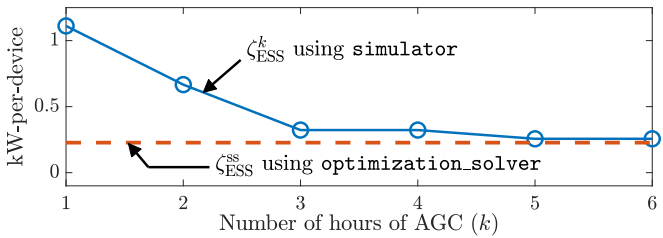}
	\vspace{-1.5em}
	\caption{The kW-per-device obtained using the \texttt{simulator} and the steady state flexibility from the macromodel are shown in this plot. As $k$ increases, $\zeta_{\text{ESS}}^{k}$ approaches $\zeta_{\text{ESS}}^{\text{ss}}$.}
	\label{fig::kW_per_ESS_multiple_hrs}
	\vspace{-1.5em}
\end{figure}

The $2$-hour kW-per-device ($\zeta_{\text{ESS}}^{2}$) obtained from Algorithm~\ref{alg:procedure_Nmin_der} resulted in the minimum number of devices required to satisfy the stopping criterion~(\ref{eq:stopping_criterion}) to be $1,500$ that reduces the kW-per-device to $\zeta_{\text{ESS}}^{2} = 0.67$kW. Similarly, extending the time period to $3$-hours further reduces kW-per-device to $\zeta_{\text{ESS}}^{3} = 0.32$kW. In Fig.~\ref{fig::kW_per_ESS_multiple_hrs}, $\zeta_{\text{ESS}}^{k}$ has been plotted for $k = 1, \dots, 6$ and shows that the flexibility reduces as the number of hours increases. This is because of the non-zero mean of the AGC that causes the fleet to either charge or discharge on average. As a result, ESS are unable to maintain the \emph{Precision} score greater than the desired $70\%$. 
Furthermore, it should be noted that in Fig.~\ref{fig::kW_per_ESS_multiple_hrs}, the kW-per-device settles to a value of about $\zeta_{\text{ESS}}^{k} = 0.26$kW after $5$ hours indicating that the flexibility approaches steady state for $k > 5$. This makes sense because the average power of the AGC signal remains approximately the same as the number of hours increases. Here, the average power of the $i$-th AGC signal $\vec{P}_{\text{agc},i}^k$ of length $K$ is defined as, 
\begin{align}
    P_{\text{avg},i}^{k} = \lim_{K \rightarrow{\infty}} \left ( \frac{1}{2K + 1}\right ) \sum_{j = -K}^K | P_{\text{agc},i}^k[j]|^2.
\end{align}
Therefore, the kW-per-device ($\zeta_{\text{ESS}}^{k}$) for $k>5$ can be considered as the steady-state value of flexibility and is denoted by $\zeta_{\text{ESS}}^{\text{ss}}$. Also, note that if $P^{k}_{\text{avg},i} > 0$ is the average power over $k$-hours, then a signal with a constant value equal to the square root of $(P^{k}_{\text{avg},i})$ over $k$-hours the same power as $\vec{P}_{\text{agc},i}^k$.

To obtain the steady-state kW-per-device using macromodel, the nominal power consumption for PEM is used which is defined as the minimum constant power signal for which QoS is sufficiently satisfied and is obtained by solving the following optimization problem~\cite{Duffaut-et-al_2019a},
\begin{subequations}   \label{eq:BetaNom}
	\begin{align}
	\min_{\beta_{\text{chg}},\beta_{\text{dis}} \in [0,1]} &\;\; h(q^\ast)  \;\;\;\; \mbox{subject to} \label{eq:MinPowerBetaNom}\\
	q^\ast   &= f(\beta, \beta^-, q^\ast), \label{eq:InvDistrBetaNom}\\
	(q^\ast)^\top \, q_{\text{v}}  &\ge x_{\text{set}}. \label{eq:SetPointEqBetaNom}
	\end{align}
\end{subequations}
By modifying the objective function to $\left (h(q^\ast) - \sqrt{P^{k}_{\text{avg},i}} \right)^2$, the non-convex optimization problem~(\ref{eq:BetaNom}) provides $P_{\text{dem}} = h(q^\ast)$ which has the same average power as $P^{k}_{\text{avg},i}$. The constraint~(\ref{eq:SetPointEqBetaNom}) ensures that the average SoC of the fleet is higher than the desired set-point to guarantee QoS, where $q_{\text{v}} \in \mathbb{R}^{4n_{\text{b}}}$ is the vector of SoC corresponding to the state vector $q$. The stopping criterion in Algorithm~\ref{alg:procedure_Nmin_der} is modified to, 
\begin{align}
    \left| P_{\text{dem}} - P^{{\text{avg}}}_{\text{agc}, i}\right| \le \epsilon_{\text{des}} \; \forall \, i = 1, \dots, m  \label{eq:stopping_criterion_power}
\end{align}
and \texttt{optimization\_solver} is used instead of \texttt{simulator} to get $\zeta_{\text{ESS}}^{\text{ss}}$. For the case of ESS, this value comes out to be $\zeta_{\text{ESS}}^{\text{ss}} = 0.23$kW which is in agreement with the long-term flexibility obtained earlier.

\section{Conclusion} \label{sec::conclusion}
This paper presented a systematic procedure to obtain flexibility in terms of minimum number of DERs, operating under a centralized and PEM coordinator, needed to track an AGC signal scaled by $\pm1$MW based on PJM's performance metrics. Flexibility is then obtained for EWHs and ESSs which is converted to a more intuitive quantity called kW-per-device. Furthermore, simulations indicate that the flexibility of a mixture of different type DERs may be obtained as a convex combination of their individual flexibility. Finally, it is observed that the flexibility over multiple hours decreases but settles to a steady state value that can be obtained using the steady state analysis of PEM state bin transition model or macromodel. Future work will focus on real-world validation of the proposed kW-per-device flexibility in the field for PEM using the setup presented in~\cite{Desrochers:2018wp}. Moreover, the historical AGC data will be analyzed to identify salient features, other than the mean values, necessary to capture flexibility. Finally, work is ongoing to obtain power and energy bounds from kW-per-device in Fig.~\ref{fig::kW_per_ESS_multiple_hrs} and will be used to study the ``trackability'' of fleets~\cite{DallAnese:CDC19}.


\end{document}